\documentclass[twocolumn,secnumarabic,amssymb, nobibnotes, aps, prl, superscriptaddress]{revtex4-2}

\usepackage{graphicx}% Include figure files
\usepackage{dcolumn}% Align table columns on decimal point
\usepackage{bm}% bold math
\usepackage{amsmath}
\usepackage{enumerate}
\usepackage{setspace}
\usepackage{dsfont}
\usepackage{subfigure}
\usepackage{multirow}
\usepackage{indentfirst} %段落缩进值宏包
\usepackage{mathrsfs}%花体字
\usepackage{color}%字体颜色
\usepackage{lineno}
\usepackage{makecell}   %表格
\usepackage{multirow}	%表格
\usepackage{upgreek}
\usepackage{threeparttable}  %表格注释
\usepackage[pdfstartview=FitH,
CJKbookmarks=true,
bookmarksnumbered=true,
bookmarksopen=true,
colorlinks, 
pdfborder=001,
linkcolor=blue,
anchorcolor=blue,
citecolor=blue
]{hyperref}

\begin{document}

\preprint{APS/123-QED}

\title{Inline-Amplification-Free Time Transfer Utilizing \\Waveform-Resolved Single-Photon Detection}% Force line breaks with \\

\author{Yufei Zhang}
\affiliation{State Key Laboratory of Advanced Optical Communication Systems and Networks, School of Electronics, and Center for Quantum Information Technology, Peking University, Beijing 100871, China}

\author{Ziyang Chen}
 \email{chenziyang@pku.edu.cn}
\affiliation{State Key Laboratory of Advanced Optical Communication Systems and Networks, School of Electronics, and Center for Quantum Information Technology, Peking University, Beijing 100871, China}

\author{Bin Luo}
\affiliation{State Key Laboratory of Information Photonics and Optical Communications, Beijing University of Posts and Telecommunications, Beijing 100876, China}

\author{Hong Guo}
 \email{hongguo@pku.edu.cn}
\affiliation{State Key Laboratory of Advanced Optical Communication Systems and Networks, School of Electronics, and Center for Quantum Information Technology, Peking University, Beijing 100871, China}

\date{\today}% It is always \today, today,
             %  but any date may be explicitly specified

\begin{abstract}
High-precision time transfer over a long haul of fiber plays a significant role in many fields. The core method, namely cascading relay nodes for the compensation of signal attenuation and dispersion, is however insufficient to deal with crucial point-to-point transfer scenarios, such as harsh environments with extremely deficient infrastructure and emergency conditions. In long-distance signal transmission without any inline amplifiers, the high loss of the optical fiber link becomes the primary limiting factor, and direct use of traditional photodetectors at the receiving end will bring about a significant drop in the stability of detected signals. Here we propose a waveform-resolved single photon detection technique and experimentally perform tomography on the weak transferred signal with an average photon number of just 0.617 per pulse. By adopting this technique, we achieve the time deviation of 95.68 ps and 192.58 ps at 200 km and 300 km respectively at an averaging time of 1 s, overcoming the technical lower bound induced by traditional photodetectors. This work lays the foundation for through-type time transfer with high precision in those significant inline-amplification-free scenarios.
\end{abstract}

%\keywords{Suggested keywords}%Use showkeys class option if keyword
                              %display desired
\maketitle

%\tableofcontents

%\section{\label{sec:level1}Introduction}
High-precision long-distance time and frequency transfer is indispensable in fields such as navigation~\cite{Jaduszliwer2021, Neil2020, Lewandowski1999}, geodesy~\cite{McGrew2018, Lisdat2016}, fundamental physics~\cite{Takamoto2020, Lisdat2016} and clock comparisons~\cite{Caldwell2023, Gozzard2022, Schioppo2022, Beloy2021, Dix2021, Oelker2019, Hyun2019, Yamaguchi2011, Ludlow2008}. Optical fiber, as an infrastructure with long transmission distances and wide coverage, is highly suitable as a medium for long-distance time and frequency transfer~\cite{Yu2024, chen2024, Gao2023, Chen2023, Shen2022, Zang2021, Zuo2021, Xue2021, Hou2019, Droste2014, Wu2014, Lopez2013, Sliwczynski2013, Predehl2012, Newbury2007}. In existing technical solutions, long-distance time and frequency transfer can be relayed through optical relays~\cite{Yu2024, Gao2023, Zang2021, Sliwczynski2013} or electrical relays~\cite{Zhang2022, Xue2021}. However, in certain special and critical application scenarios, it is often difficult to establish effective relay nodes. For example, in areas with extremely harsh environments, such as remote areas with very underdeveloped infrastructure, the deployment of relay nodes faces challenges. In scenarios lacking long-distance signal regeneration technology, such as deserts, plateaus, polar regions, or undersea~\cite{Stafford1995}, optical cable laying can only support point-to-point transmission. In burst scenarios like war, natural disasters, or emergency communications, the flexibility of relay deployment is greatly reduced. These scenarios urgently demand high-precision, relay-free time and frequency transfer technology.

In long-distance signal transmission without inline amplification, compared to the effects of dispersion, nonlinearities, and other factors, the high loss of the optical fiber link becomes the primary limiting factor~\cite{Agrawal5}. For example, even under ideal fiber attenuation, the typical loss for a 300-km optical fiber link would reach 60 dB. To address the high loss in long-distance signal transmission, there are two typical methods in classical optical communication for extending the transmission distance without relay. One approach involves placing amplifiers after the transmitter and before the receiver, enabling signal transmission over distances exceeding 500 km at a rate of 2.5 Gb/s~\cite{Hansen1995}. The other approach employs novel detection methods, such as single-photon detectors, which utilize the avalanche effect to amplify weak photocurrents, thereby enabling the detection of weak signals~\cite{Zhou2023, Chen2022, Wang2022, Chen2021, Lucamarini2018, Takesue2007}.

However, in high-precision time and frequency transfer, both of these approaches present theoretical and experimental challenges. At the transmitting end, the aforementioned transmission schemes require ultra-high input power, posing significant experimental limitations such as the use of broad-linewidth laser to increase the stimulated Brillouin scattering (SBS) threshold~\cite{Agrawal2, Goodno2019, Liu2017, Anderson2014, Hansen1995}. At the receiving end, on the one hand, the limited gain-bandwidth product (GBP)~\cite{He2021, Sorger2020} necessitates a trade-off between achieving high gain and maintaining a sufficiently large detection bandwidth, meaning that a balance must be achieved between receiving enough energy in the signal and preserving the signal's rise time, phase, and other information without severe distortion. This leads to a technical lower bound on the stability of the transmitted signal; under typical gain-bandwidth products, the second stability of time transfer measurements over a 300-km scale deteriorates to the order of 10 nanoseconds~\cite{zhang2024}. On the other hand, the use of single-photon detectors in the above schemes only allows for judging the high and low levels of the signal, corresponding to the presence or absence of photons, and loses the temporal information of pulse details. In high-precision time and frequency transfer, however, it is necessary to detect the waveform of the transmitted signal to obtain key information for stability evaluation and time synchronization, such as phase jitter and rise time position. Therefore, solutions for the relay-free scenarios in classical optical communication still fall short of the requirements for high-precision time and frequency transfer.

In this Letter, we present the waveform-resolved single photon detection technology at the receiving end to achieve long-distance high-precision time and frequency transfer without any inline amplification. By dynamically scanning the photon energy integration interval, we successfully recover the waveform of extremely weak light pulses, as well as critical timing information such as the pulse rise time. Simultaneously, at the transmitting end, we employ high-speed phase modulation of a continuous light source to increase the SBS threshold, thereby boosting the input fiber power. Ultimately, after transmitting through a 300-km standard optical fiber link without any inline amplification, we measure extremely weak light with an average photon number of only 0.617 per pulse, achieving the second instability of 192.58 ps. Under the new single-photon detection technology, we not only reconstruct extremely weak waveforms at the remote end but also achieve a time deviation (TDEV) metric nearly two orders of magnitude better than the technical lower bound provided by traditional schemes. To our knowledge, this work represents the longest-distance high-precision time and frequency transfer reported to date under full-loss conditions without any amplification in the fiber link. Moreover, the waveform-resolved single-photon detection technology we propose enables the recovery of extremely weak time-frequency signal waveforms and the extraction of significant timing information. We believe this technology is versatile and will generate widespread interest among researchers in fields such as time and frequency transfer, signal processing, and communication.

%\section{\label{sec:level2}Experimental Setup and Methods}

\begin{figure*}[t]
\includegraphics[width = 1.0\textwidth]{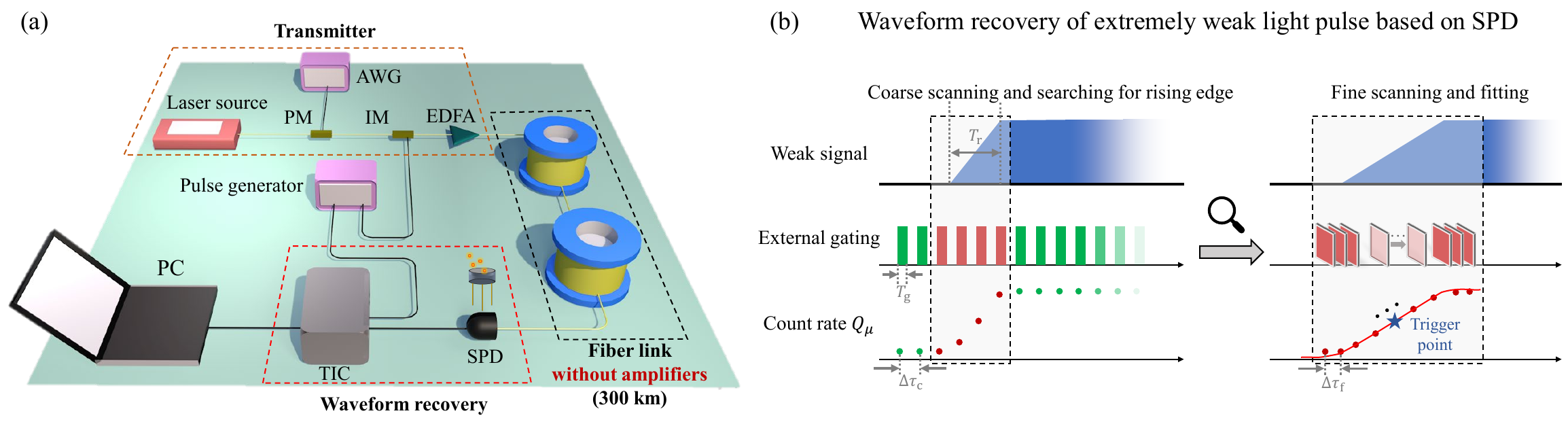}
\caption{The long-distance inline-amplifier-free time transfer protocol based on single-photon detection technology, where (a) the experimental setup and (b) schematic diagram of waveform recovery process are depicted respectively. PM: phase modulator, IM: intensity modulator, AWG: arbitrary waveform generator, EDFA: erbium-doped fiber amplifier, SPD: single-photon detector, TIC: time interval counter.}
\label{diagram}
\end{figure*}

Figure~\ref{diagram}(a) shows a schematic diagram of the long-distance inline-amplifier-free time transfer protocol based on waveform- resolved single-photon detection technology. It primarily consists of three components: the transmitter, channel, and waveform recovery, which are detailed in the following text respectively.

\textbf{Transmitter}. First, the narrow linewidth continuous wave (cw) light output from the laser source, with a linewidth of $\nu_0$, undergoes phase modulation to increase the SBS threshold. This reduces the backward Brillouin scattered light and increases the optical power input into the fiber to extend the transmission distance. Let the linewidth of the laser after modulation be $\nu_p$, then the threshold for stimulated Brillouin scattering is given by~\cite{Anderson2014, Aoki1988}
\begin{equation}
    P_\text{th} = \frac{21A_\text{e}}{g_\text{b}L_\text{e}}(1+\frac{\nu_\text{p}}{\nu_\text{b}}),
    \label{Eq1}
\end{equation}
where $A_\text{e}$ represents the effective mode field area of the optical fiber, $g_\text{b}$ is the Brillouin gain coefficient, $L_\text{e} = \frac{1 - \exp(-\alpha L)}{\alpha}$ is the effective length of the optical fiber, and $\nu_\text{b}$ is the Brillouin gain bandwidth. The clock signal at frequency $f_\text{r}$ is loaded through an intensity modulator onto the cw optical carrier with increased bandwidth. After being amplified by an erbium-doped fiber amplifier (EDFA) at the source end, the pulse sequence carrying the clock signal is injected into the optical fiber link.

\textbf{Channel}. A single-span optical fiber link with the length of $L$[km] is deployed, which has no relay nodes or distributed amplification. The attenuation coefficient of optical fiber is $\alpha~[\text{km}^{-1}]$.

\textbf{Waveform recovery}. The extremely weak optical signal, after being transmitted over a long distance, is input into a single-photon detector where its waveform is recovered. The detection efficiency of the single-photon detector is $\eta$, the dead time is $t_\text{d}$, and the dark count rate is $P_\text{dark}$. As shown in Fig.~\ref{diagram}(b), we choose to operate the single-photon detector in external-gating mode, with the gating signal having the same frequency $f_\text{r}$ as the clock signal. We vary the relative time delay $\tau$ of the applied gating signal within one period to scan the time domain of the extremely weak optical pulses. The change in the single-photon detector's count rate $Q_\mu$ is detected using a frequency counter, where the subscript $\mu$ represents the average number of photons per pulse in the received extremely weak signal, satisfying the following formula~\cite{Lo2005, Hwang2003}
\begin{equation}
    Q_\mu = 2P_\text{dark}+1-e^{-\eta\mu}.
    \label{Eq2}
\end{equation}
The detailed derivation and analysis of Eq.~(\ref{Eq2}) can be found in the Supplemental Document~\cite{supplemental}.

Since the count rate $Q_\mu$ reflects the probability of the single-photon detector registering a count after energy integration over a time gate of certain width, it is proportional to the average photon number at different phases of the transmitted periodic signal. Therefore, by scanning the relative delay $\tau$ and measuring the count rate $Q_\mu$, we can obtain the relative intensity at different phases of the transmitted periodic signal. This enables a complete temporal waveform scan under weak light conditions, thereby achieving tomography of extremely weak signal waveforms. And we perform high-speed repetitive scans of the single-photon detector count rate $Q_\mu$ at each pulse delay $\tau$ and average the count rate at each point. We consider the average count rate $Q_\mu$ of the single-photon detector at each pulse time position 
$\tau$ as the scanning result for the relative intensity of the temporal pulse reconstruction under weak light conditions. By fitting this waveform, we obtain an estimate of the extremely weak signal waveform. This method aims to restore as much detail as possible, with particular focus on recovering the critical temporal information of the rising edge in this work.

%\section{\label{sec:level3}Results and Discussions}
In the experiment, we used an NKT laser (Koheras BASIK X15) as the light source, whose linewidth $\nu_0 < 0.1$ kHz. Here, the Gaussian white noise phase modulation was applied to increase the linewidth and raise the Brillouin scattering threshold. We used a phase modulator and applied Gaussian white noise with a modulation frequency of 250 MHz generated by an arbitrary waveform generator to increase the optical carrier’s bandwidth and therefore raise the SBS threshold. The experiment employed the single-mode fiber, with an effective mode field area $A_\text{e}\approx\pi\times(10~\upmu\text{m})^2$, an attenuation coefficient of approximately 0.2 dB/km, and for long distances, the effective fiber length $L_\text{e}\approx1/\alpha$, where $\alpha_{(1/\text{km})} = \alpha_{(\text{dB/km})}\times\ln 10/10$. In addition, the stimulated Brillouin scattering gain coefficient $g_\text{b}\approx 5 \times 10^{-11}~\text{m/W}$, and the typical value of the gain bandwidth $\nu_\text{b}$ is 50 MHz~\cite{Agrawal2}. Substituting these parameters into Eq.~(\ref{Eq1}) gives the stimulated Brillouin scattering threshold $P_\text{th} \approx 36~\text{mW}$, which provides the basis for coupling the light carrying clock information from the transmitter into the optical fiber link after amplification by the EDFA. Experimentally, after undergoing phase modulation, the cw light was further loaded with the clock signal, generated by the frequency source with a frequency $f_\text{r} = 1~\text{kHz}$ and a duty cycle $D = 20\%$ through intensity modulation.

\begin{figure}[t]
\includegraphics[width = 0.48\textwidth]{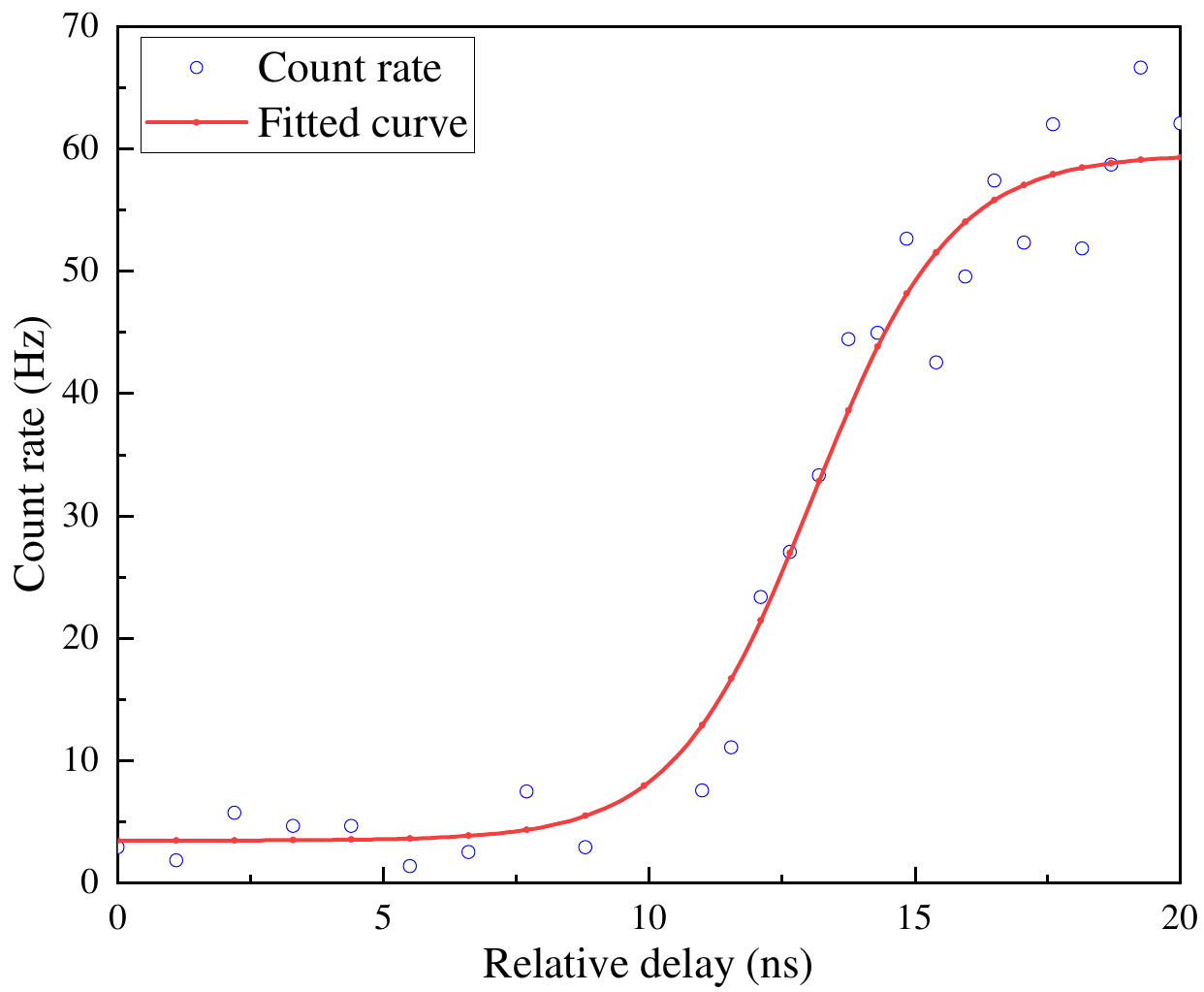}
\caption{Tomography of extremely weak optical signals after transmission through a 300-km all-loss optical fiber link and reconstruction of its rising edge.}
\label{fitting}
\end{figure}

Based on the power estimation, the pulse entering the fiber contains an average photon number on the order of $10^{13}$/pulse. After passing through a 300-km optical fiber link without inline amplification and over 60 dB of attenuation, the extremely weak optical signal carrying the clock information from the transmitter was input into the single-photon detector. Using Eq.~(\ref{Eq2}), the average photon number is calculated to be approximately $\mu = 0.617$/pulse. Figure~\ref{diagram}(b) gives more details on how to experimentally recover extremely weak signal waveforms through tomographic scanning. Firstly, we performed a wide-range tomography, covering at least one complete cycle of the extremely weak signal by adjusting the initial phase $\phi_0$ of the external gating signal. The gate width is $T_\text{g}=5~\text{ns}$, shorter than the rising edge of the pulse signal $T_\text{r}\sim20~\text{ns}$. The time resolution for the weak signal waveform is $\Delta\tau_\text{c}=\frac{\Delta\phi_0}{2\pi}\times\frac{1}{f_\text{r}}=2.78~\text{ns}$, where $\Delta\phi_0 = 0.001^\circ$ is the minimum step size of the initial phase. Secondly, after adjusting the gate to be near the rising edge of the signal, we further refined the tomography by adjusting the trigger delay of the single-photon detector with a step size of $\Delta\tau_\text{f} = 0.55~\text{ns}$ to carefully analyze the rising edge and extract the critical timing information. We collected the count rate $Q_\mu$ at each relative delay and recovered 20 sets of rising edge waveforms through second averaging. These waveforms were then fitted in the third step by exploiting the Sigmoid function
\begin{equation}
    Q_\mu=A+\frac{B}{1+\exp(-k(\tau-\tau_0))},
    \label{Eq3}
\end{equation}
where the four parameters $A$, $B$, $k$ and $\tau_0$ are determined through fitting. The base of the pulse rising edge is approximately $A$, and the top of the rising edge is approximately $A + B$. The parameters $k$ and $\tau_0$ reflect the slope of the rising edge and its central position respectively. The Sigmoid function is a well-performing function commonly used for fitting signals that distinguish between 0 and 1~\cite{Lee2018}. In Fig.~\ref{fitting}, the blue circles represent the raw data points obtained from measuring the count rate, while the red curve represents the fitted rising edge.

\begin{figure}[t]
\includegraphics[width = 0.48\textwidth]{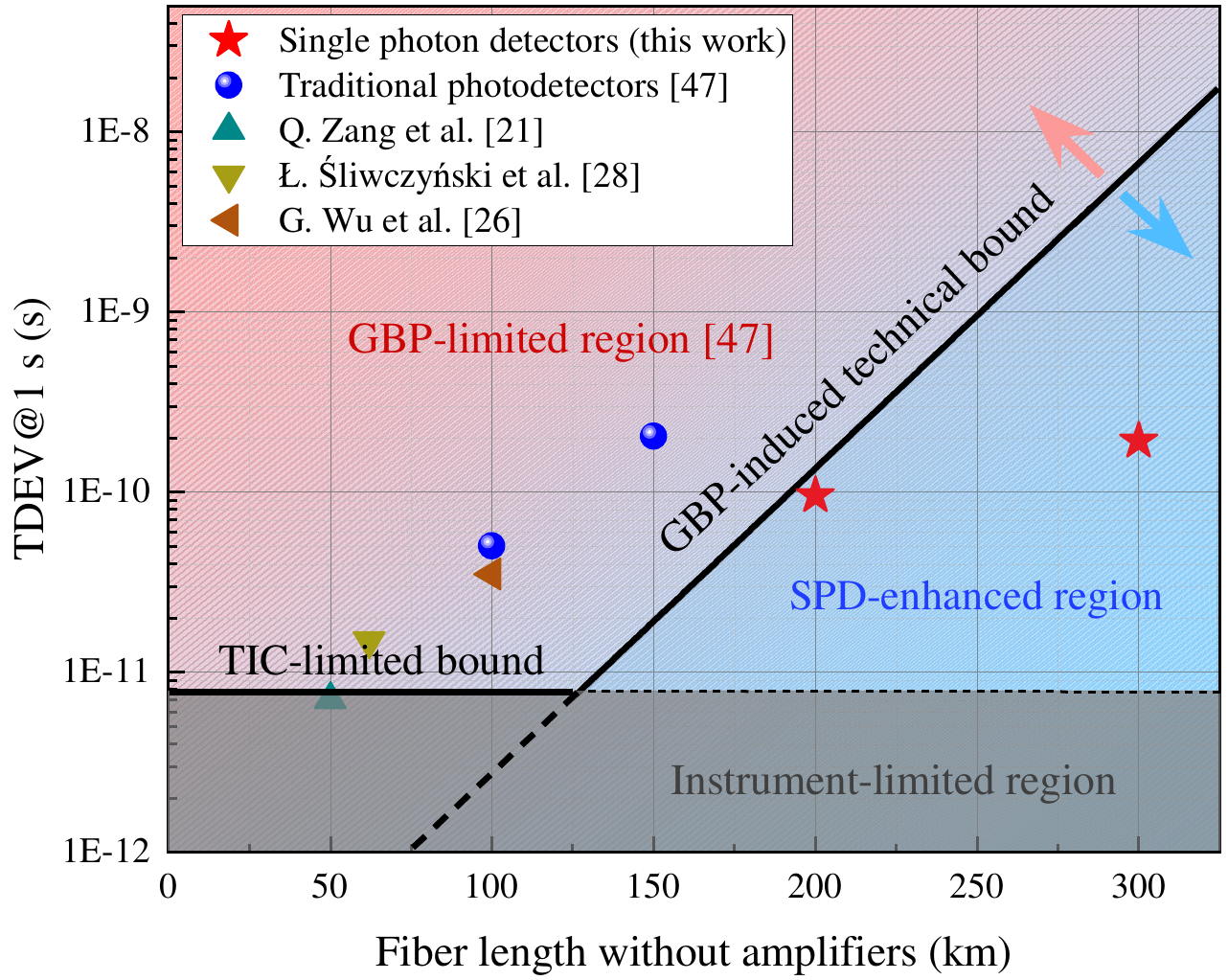}
\caption{Comparison between the waveform-resolved single-photon detection scheme proposed in this paper and the traditional photodetector-based detection scheme in time transfer without inline amplification.}
\label{results}
\end{figure}

We selected the 50\% level of the recovered rising edge as the trigger point and sequentially recorded the trigger times of the 20 reconstructed signals. Based on this, we calculated the TDEV instability. Figure~\ref{results} presents the main performance metrics achieved utilizing the waveform-resolved single-photon detection technique in this work, along with a comparison to traditional detection schemes. The black solid line in Fig.~\ref{results}, referenced from literature~\cite{zhang2024}, represents the technical lower bound for time transfer without any inline amplifiers under a 50-GHz gain-bandwidth product at the detection end within the traditional detection framework. At shorter distances, the primary limitation comes from the time interval measurement instrument, with the figure showing the background noise of the time interval counter (Agilent 53230A) used in our experiment. At longer distances, the limitation is due to the finite gain-bandwidth product at the detection end. The triangles and purple circles in the figure represent experimental results using traditional detectors as reported in the literature~\cite{Zang2021, Wu2014, Sliwczynski2013}, all of which fall above the boundary defined by the technical bound. The red pentagrams in the figure represent the main results obtained using the new single-photon detection scheme. At distances of 200 km and 300 km, time instabilities of 95.683 ps and 192.58 ps, respectively, were achieved through waveform resolution and reconstruction. As shown in Fig.~\ref{results}, the data points measured using the new single-photon detection scheme all fall below the boundary defined by the black solid line, surpassing the technical bound established under the traditional detection framework~\cite{zhang2024}. 

Moreover, by comparing the variation in time instability with distance for the traditional detection scheme represented by blue circles and the single-photon detection scheme represented by red pentagrams, it can be seen that the time signal stability degrades more slowly with distance when using the new single-photon detection technology with waveform resolution. This is because the single-photon detector integrates the energy of the weak received light signal within a fixed gate time, forming an avalanche current and outputting a count signal with a stable shape. As a result, it detects the intensity of the received light signal without being sensitive to the quality of the received waveform. Therefore, for short-distance time transfer without inline amplification, the detection scheme based on traditional photodetectors is sufficient. However, for long-distance time transfer on the order of 200 km and beyond, the traditional scheme is constrained by the technical bound and can no longer meet the high-precision time transfer requirements. In this scenario, the alternative single-photon detection scheme proposed in this paper offers an effective solution.

%\section{\label{sec:level4}Conclusions}
In summary, we proposed and demonstrated a waveform-resolved single-photon detection technique, in response to the challenge of recovering extremely weak signals in the important application scenario of long-distance time signal transfer without inline amplification. By dynamically scanning the time-domain waveform of the weak signal, this technique enables the recovery of the extremely weak light pulse waveform and the extraction of key pulse timing information, particularly the extraction of the pulse rise time. Moreover, our detection scheme operates under gated conditions, recording the count rate to characterize signal strength. This waveform tomography method can reduce the impact of the single-photon detector's time resolution by relaxing the requirement of recording individual photon arrival times.

Using this technique, we successfully performed tomography on a full-loss weak signal with an average photon number of just 0.617 per pulse, and under conditions without any inline amplification in the fiber link, we achieved TDEV of 95.68 ps at a distance of 200 km and 192.58 ps at 300 km, at an averaging time of 1 s. For even longer distance, improvements could be made by increasing the incident light linewidth to raise the Brillouin scattering threshold, enhancing single-photon detection efficiency, and reducing dark count rates. We believe that this work will not only advance research and applications in the important scenario of through-type time transfer without inline amplification, but also provide references for the recovery of extremely weak signal waveforms and the extraction of key timing information across various fields including optical communication and clock networks.

%\begin{acknowledgments}
We acknowledge funding from the National Natural Science Foundation of China (Grant No. 62201012) and the National Hi-Tech Research and Development (863) Program.
%\end{acknowledgments}

\appendix
\section{Supplement}
\setcounter{equation}{0}
\renewcommand{\theequation}{S\arabic{equation}}
This supplement provides details on the derivation and analysis of Eq.~(\ref{Eq2}) in the main text.

Let $P_n$ represent the probability of an $n$-photon signal appearing in extremely weak light after being transmitted through the optical fiber link, and $Y_n$ represent the probability that the single-photon detector responds when an $n$-photon signal is received. Therefore, the probability that an $n$-photon signal is detected and the single-photon detector responds is $Q_n=P_nY_n$. Furthermore, considering the possibility of multiple photons being received, the count rate of the single-photon detector is given by
\begin{equation}
    Q_\mu=\sum_{n=0}^\infty Q_n=\sum_{n=0}^\infty P_nY_n,
\label{S1}
\end{equation}
where the subscript $\mu$ represents the mean photon number in the received signal. Here, we assume that the photon number distribution follows a Poisson distribution, i.e.,
\begin{equation}
    P_n=e^{-\mu}\frac{\mu^n}{n!}.
\label{S2}
\end{equation}

In the experiment, a non-photon-number-resolving (NPNR) detector is used. As long as one photon triggers the avalanche effect, the single-photon detector will register a count. After comprehensively considering dark counts and multiphoton processes, the final expression for the count rate of the single-photon detector at the detection end can be given as follows
\begin{equation}
    Q_\mu=Y_0+1-e^{-\eta\mu},
\label{S3}
\end{equation}
where $Y_0\approx2P_\text{dark}$ represents the response when there are zero photons, and $\eta$ is the detection efficiency of the single-photon detector. In the experiment, the single-photon detector measures an extremely weak signal after passing through a 300-km all-loss optical fiber link without any inline amplification, and yields a count rate of approximately $Q_\mu=0.06~\text{count/pulse}$. The dark count rate of the single-photon detector used in the experiment is approximately $P_\text{dark}=6\times10^{-5}~\text{count/pulse}$, and the detection efficiency is $\eta=10\%$. Substituting these parameters into Eq.~(\ref{S3}), the average photon number is calculated to be $\mu=0.617/\text{pulse}$, indicating that after transmission through a 300-km optical fiber link without any inline amplification, the extremely weak signal contains an average photon number of less than 1/pulse.

% The \nocite command causes all entries in a bibliography to be printed out
% whether or not they are actually referenced in the text. This is appropriate
% for the sample file to show the different styles of references, but authors
% most likely will not want to use it.
%\nocite{*}

%\bibliography{apssamp}% Produces the bibliography via BibTeX.
%apsrev4-2.bst 2019-01-14 (MD) hand-edited version of apsrev4-1.bst
%Control: key (0)
%Control: author (8) initials jnrlst
%Control: editor formatted (1) identically to author
%Control: production of article title (0) allowed
%Control: page (0) single
%Control: year (1) truncated
%Control: production of eprint (0) enabled
%

\end{document}